\documentclass[english,showpacs,floatfix,11pt]{revtex4}
\usepackage[dvips]{graphicx}
\usepackage{longtable}
\usepackage{amsmath,amssymb}
\usepackage{dcolumn}
\usepackage{latexsym}
\usepackage{babel}
\usepackage[latin1]{inputenc}
\usepackage{color}
\usepackage[colorlinks]{hyperref}
\def\al{\alpha}
\def\kp{\kappa}

\def\nb{\nabla}
\def\pa{\partial}
\def\vf{\varphi}

\def\ga{\gamma}
\def\be{\beta}
\def\ck{\check}

\def\la{\lambda}

\def\th{\theta}
\def\sg{\sigma}

\def\nn{\nonumber}
\def\diag{\mbox {diag}}

\def\l{\left}
\def\r{\right}

\begin{document}
\title{\bf Integrable conditions  for  Dirac Equation and Schr\"odinger equation}
\author{Ying-Qiu Gu}
\email{yqgu@fudan.edu.cn} \affiliation{School of Mathematical
Science, Fudan University, Shanghai 200433, China} 
\date{19th May 2017}

\begin{abstract}
By constructing the commutative operators chain, we derive the
integrable conditions for solving the eigenfunctions of Dirac
equation and Schr\"odinger equation. These commutative relations
correspond to the intrinsic symmetry of the physical system, which
are equivalent to the original partial differential equation can be
solved by separation of variables. Detailed calculation shows that,
only a few cases can be completely solved by separation of
variables. In general cases, we have to solve the Dirac equation and
Schr\"odinger equation by effective perturbation or approximation
methods, especially in the cases including nonlinear potential or
self interactive potentials. \vskip3mm \noindent{Keywords:} {\sl
eigenfunction, commutative relation, separation of variables,
nonlinear Dirac equation} 

\end{abstract}
\maketitle

\section{Introduction}
\setcounter{equation}{0}

All fermions have spin-$\frac 1 2$, which are naturally described by
spinors. Many physicists such as H. Weyl, W. Heisenberg, once
proposed using nonlinear spinor equations to establish unified field
theory for elementary particles\cite{2,3}. Some rigorous solutions
for the simplest dark nonlinear spinor models were obtained and
analyzed\cite{4}-\cite{11}. The nonlinear spinor coupling with
self-electromagnetic field was calculated in \cite{12}-\cite{15},
but only roughly approximate results were obtained due to the
complexity of the interaction. In contrast, for the linear
Schr\"odinger equation, Pauli equation and Dirac equation coupling
with external potential $V(r)$ and $\vec A(r,\th)$, there are a lot
of rigorous eigen solutions obtained by separation of
variables\cite{17}-\cite{27}.

In the resolution of eigenfunctions to the linear Dirac equation,
we find that the commutative relations\cite{17}
\begin{eqnarray}
[\hat H, \hat J_z]=0,\qquad [\hat H,\hat K]=0,\qquad [\hat K,\hat
J_z]=0 \label{comm}\end{eqnarray} play an important role, where
the operators $(\hat J_z, \hat K, \hat H)$ stand for angular
momentum, spin-orbit coupling and total energy operators
respectively. They are good quantum numbers guaranteeing the
common eigenfunctions exist. Noticing the definition of operators
chain
\begin{eqnarray}
\hat J_z=J_z(\pa_\vf),\qquad \hat K=K(\pa_\vf,\pa_\th), \qquad \hat
H=H(\pa_\vf,\pa_\th,\pa_r), \label{chain}\end{eqnarray} which enable
us to solve the eigen solutions by separation of variables, and then
the original problem reduces to ordinary differential equations.

Denote the Minkowski metric by $\eta_{\mu\nu}={\rm
diag}[1,-1,-1,-1]$, Pauli matrices by
\begin{equation}
 {\vec\sg}=(\sg^{j})= \left\{\left(\begin{matrix}
 0 & 1 \cr 1 & 0\end{matrix}\right),~\left(\begin{matrix}
 0 & -i \cr i & 0\end{matrix}\right),~\left(\begin{matrix}
 1 & 0 \cr 0 & -1\end{matrix}\right)\right\}
 .\label{1.1}\end{equation}
Define $4\times4$ Hermitian matrices as follows
\begin{equation}\al^\mu=\left\{\left ( \begin{array}{ll} I & ~0 \\
0 & I \end{array} \right),\left (\begin{array}{ll} 0 & \vec\sg \\
\vec\sg & 0 \end{array}
\right)\right\},~ \ga =\left ( \begin{array}{ll} I & ~0 \\
0 & -I \end{array} \right),
~ \be=\left (\begin{array}{ll} 0 & -iI \\
iI & ~~0 \end{array} \right).\label{1.2}
\end{equation}
In this paper, we adopt the Hermitian matrices (\ref{1.2}) instead
of Dirac matrices $\ga^\mu$, because this form is more convenient
for calculation.

Our basic problem is to examine a nonlinear spinor field $\phi$
moving in 4-vector potential $A^{\mu}$. The corresponding Lagrangian
is generally given by\cite{13}
\begin{equation}
{\cal L} =\phi^+[\al^\mu (\hbar i\pa_\mu-eA_\mu) -\mu c
\ga]\phi+F(\ck\ga,\ck\be), \label{1.3}\end{equation} where $\mu>0$
is a constant mass, $F$ is the nonlinear coupling potential, which
is the polynomial of $\ck\ga$ and $\ck\be$ defined by
\begin{equation}
\ck\ga=\phi^{+}\ga\phi, \qquad \ck\be=\phi^{+}\be\phi.
\label{1.4}\end{equation} It is well known that $\ck\ga$ is a
true-scalar and $\ck\be$ a pseudo-scalar. The variation of
(\ref{1.3}) with respect to $\phi^+$ gives the dynamic equation
\begin{equation}
\hbar i \pa_0\phi=\hat H\phi,  \quad \hat H\equiv
\vec\al\cdot(-\hbar i \nb -e\vec A)+eA_0+(\mu c
-F_\ga)\ga-F_\be\be,\label{1.5}
\end{equation}
where $F_\ga=\frac{\pa F}{\pa\ck\ga}, F_\be=\frac{\pa
F}{\pa\ck\be}$.

Let coordinate $x^3=z$ along the direction of magnetic field $\vec
B$, then we locally have\cite{9,10,13}
\begin{equation} \vec A=A(r,\th)(-\sin\vf,\cos\vf,0),\label{1.6}
\end{equation}
which satisfies the Coulomb gauge $\nb \cdot \vec A=0$. In the
spherical coordinate system $(r, \th, \vf)$, we have
\begin{equation}
\vec \sg\cdot \nb= \sg_r\pa_r + \frac 1 r \sg_\th\pa_\th+\frac 1
{r\sin\th}\sg_\vf\pa_\vf,\label{1.7}
\end{equation}
where $(\sg_r, \sg_\th, \sg_\vf)$ is given by
\begin{equation}
\left\{\left(\begin{matrix}\cos\th & \sin \th e^{-\vf i}\cr\sin\th
e^{\vf i}
 & -\cos\th\end{matrix}\right),
\left(\begin{matrix}-\sin\th & \cos\th e^{-\vf i}\cr\cos\th e^{\vf
i} & \sin\th\end{matrix}\right), \left(\begin{matrix}0 &-i e^{-\vf
i}\cr i e^{\vf i} & 0\end{matrix}\right)\right\}.\label{1.8}
\end{equation}
Let $\hat J$ be the angular momentum operator for the spinor field
\begin{equation}
\hat J=\vec r \times \hat p+\frac 1 2 \hbar \vec S,\quad \hat p=
-\hbar i\nb,\quad \vec S=\diag (\vec\sg,\vec\sg). \label{1.9}
\end{equation}
Then the angular momentum $\hat J_3=\hat J_z$ commutates with the
nonlinear Hamilton operator (\ref{1.5}), and the eigenfunctions of
$\hat J_z=-\hbar i \pa_\vf+\frac 1 2 \hbar S_3$ are equivalent to
the following form
\begin{equation}\phi=(u_1,u_2e^{\vf i},-iv_1,-iv_2e^{\vf i})^T \exp\left(\kp \vf
i-\frac{mc^2} \hbar ti\right)\label{1.10}\end{equation} with
$(\kp=0,\pm 1,\pm 2,\cdots)$, where $u_k, v_k(k=1,2)$ are real
functions of $(r, \th)$ but independent on $\vf$ and $t$, the
index $T$ stands for transposed matrix.

\section{Commutative algebras for Dirac equation}
\setcounter{equation}{0}

In spherical coordinate system, by straightforward calculation, we
have the following explicit operator relations\cite{17}
\begin{eqnarray}
\pa_x&=&\sin\th\cos\vf\pa_r+\frac 1 r \cos\th\cos\vf\pa_\th-\frac
1 {r\sin\th}\sin\vf\pa_\vf,\label{2.1}\\
\pa_y&=&\sin\th\sin\vf\pa_r+\frac 1 r \cos\th\sin\vf\pa_\th+\frac
1 {r\sin\th}\cos\vf\pa_\vf, \label{2.2}\\
\pa_z&=& \cos\th\pa_r-\frac 1 {r\sin\th}\pa_\vf ,\label{2.3}
\end{eqnarray}
For orbit angular momentum operator $\hat L=\vec r\times \hat p$,
we have
\begin{eqnarray}
\hat L_x&=&\hbar i \l(\sin\vf\pa_\th+\cot\th\cos\vf\pa_\vf\r), \label{2.4}\\
\hat L_y&=&\hbar i \l(-\cos\vf\pa_\th+\cot\th\sin\vf\pa_\vf\r), \label{2.5}\\
\hat L_z&=& -\hbar i\pa_\vf ,\label{2.6}\\
\hat L^2 &=& -\hbar^2 (\pa_\th^2+\cot\th\pa_\th+\frac
1{\sin\th^2}\pa_\vf^2 ). \label{2.7}
\end{eqnarray}
For the nonlinear equation (\ref{1.5}), detailed calculation shows
that
\begin{eqnarray}
[\hat H,\hat J_z] = 0. \label{2.10}
\end{eqnarray}
(\ref{2.10}) shows $\hat J_z$ is still a good quantum number for the
nonlinear Dirac equation. But $[\hat H,J_x]\ne0, [\hat H,J_x]\ne0$.
Define the spin-orbit coupling operator by
\begin{eqnarray}
\hat K\equiv \ga(\hat L \cdot\vec S+\hbar)= \hbar\ga-\hbar i
(K_\th\pa_\th+K_\vf\pa_\vf), \label{2.18}
\end{eqnarray}
where $K_\th$ and $K_\vf$ defined by
\begin{eqnarray}
K_\th &=&\diag \l[\l( \begin {array}{cc} 0 & -ie^{-\vf i} \\
\noalign{\medskip} i e^{\vf i} & 0 \end {array} \r),\l( \begin
{array}{cc} 0 & i e^{-\vf i} \\\noalign{\medskip} -i e^{\vf i} & 0
\end {array} \r)\r] , \label{2.19}\\
K_\vf &= &\diag \l[\l( \begin {array}{cc} 1 & -\cot\th e^{-\vf i} \\
\noalign{\medskip} -\cot\th e^{\vf i} & -1 \end {array} \r),\l(
\begin {array}{cc} -1 & \cot\th e^{-\vf i} \\\noalign{\medskip} \cot\th e^{\vf i}
& 1 \end {array} \r)\r], \label{2.20}
\end{eqnarray}
then for (\ref{1.5}), we can check
\begin{eqnarray}
[\hat H,\hat K]=K_V+K_A+K_\be, \label{2.22}
\end{eqnarray}
in which
\begin{eqnarray}
K_V &=&-\hbar i(e\pa_\th V  -\pa_\th F_\ga\ga)\ga(\sin\vf S_1-\cos\vf S_2),\label{2.22.1}\\
K_A &=&- \hbar e \l(A\ga(\sin\vf \al^1-\cos\vf \al^2)-(\pa_\th
A+\cot\th A+2A\pa_\th)\be \r), \label{2.23} \\
K_\be &=& 2\hbar \ga\be F_\be+\hbar(\pa_\th
F_\be+2F_\be\pa_\th)(\sin\vf\al^1-\cos\vf \al^2)+ \nn\\
&~&2\hbar F_\be \l( \cot\th(\cos\vf
\al^1+\sin\vf\al^2)-\al^3\r)\pa_\vf . \label{2.24}
\end{eqnarray}
(\ref{2.22})-(\ref{2.24}) reflect the influence of parameters on the
integrability.

\section{Conditions for Separation of Variables}
\setcounter{equation}{0}

For the nonlinear Hamiltonian (\ref{1.5}), the commutative relation
(\ref{2.10}) shows $\hat J_z$ is still a good quantum number, but
(\ref{2.22}) shows $\hat K$ is not. In what follows, we look for a
new operator $\hat T$, such that the systems (\ref{1.5}) can be
solved by separation of variables. This is equivalent to the
existence of operator $\hat T=T(\pa_\th,\pa_\vf)$ or $\hat
T'=T'(\pa_r,\pa_\vf)$, which satisfies commutative relations
\begin{eqnarray}
[\hat J_z, \hat T] = 0,\qquad [\hat H,\hat T]=0. \label{3.1}
\end{eqnarray}
We only consider the following first order
operator
\begin{eqnarray} \hat T = T_0-\hbar i(T_\th
\pa_\th+T_\vf\pa_\vf),\label{3.2}
\end{eqnarray}
where $(T_0, T_\th, T_\vf)$ are all Hermitian matrices, and their
components are smooth functions of $(r,\th,\vf)$. For Dirac
equation, it is enough to look for the linear operators similar to
(\ref{3.2}). But for the Schr\"odinger and Pauli equations, we
should consider the second order operator similar to $\hat L^2$ as
shown below. Different from $\hat p$ and $\hat J$, the operators
constructed from the procedure may have not manifest physical
meanings.

According condition $[\hat J_z, \hat T] = 0$, we can solve
\begin{eqnarray}
T_0 =  \left( \begin {array}{cccc}  P_{11} & P_{12}  \cot\th
e^{-i\phi}&
 P_{13} & P_{14} \cot\th  e^{-i\phi}
\\\noalign{\medskip} P_{21} \cot \th  e^{i\phi} & P_{22}
 &  P_{23} \cot
\th e^{i\phi} &P_{24} \\\noalign{\medskip} P_{31} & P_{32}\cot\th
e^{-i\phi} &P_{33}
 & P_{34}\cot\th  e^{-i\phi}
\\\noalign{\medskip}P_{41}\cot \th  e^{i\phi} & P_{42}
& P_{43}\cot\th e^{i\phi}& P_{44} \end {array} \right),
\end{eqnarray}
\begin{eqnarray}
T_\th = \left( \begin {array}{cccc}  M_{11} & M_{12} e^{-i\phi}&
 M_{13} & M_{14}    e^{-i\phi}
\\\noalign{\medskip} M_{21}    e^{i\phi} & M_{22}
 &  M_{23}   e^{i\phi} &M_{24} \\\noalign{\medskip} M_{31} & M_{32}
e^{-i\phi} &M_{33}
 & M_{34}   e^{-i\phi}
\\\noalign{\medskip}M_{41}   e^{i\phi} & M_{42}
& M_{43}  e^{i\phi}& M_{44} \end {array} \right),
\end{eqnarray}
\begin{eqnarray}
T_\vf = \left(
\begin {array}{cccc}  N_{11} & N_{12}  \cot\th e^{-i\phi}&
 N_{13} & N_{14} \cot\th  e^{-i\phi}
\\\noalign{\medskip} N_{21} \cot \th  e^{i\phi} & N_{22}
 &  N_{23} \cot
\th e^{i\phi} &N_{24} \\\noalign{\medskip} N_{31} & N_{32}\cot\th
e^{-i\phi} &N_{33}
 & N_{34}\cot\th  e^{-i\phi}
\\\noalign{\medskip}N_{41}\cot \th  e^{i\phi} & N_{42}
& N_{43}\cot\th e^{i\phi}& N_{44} \end {array} \right),
\end{eqnarray}
where $(P_{kl}, M_{kl}, N_{kl})$ are Hermitian matrices, and their
components are functions of $(r,\th)$, the factor `$\cot\th$' is
introduced for convenience of the following calculation.

The relation $[\hat H, \hat T]$ can be expressed as
\begin{eqnarray}
[\hat H, \hat T] &=&
H_{\th\th}\pa_\th^2+H_{r\th}\pa_{r\th}+\l(H_{\th}+H_{\th\vf}\pa_\vf\r)\pa_\th+\l(H_{r}+H_{r\vf
}\pa_\vf
\r)\pa_r\nn\\
&~& +\l(H_0+H_{\vf}\pa_\vf+H_{\vf\vf}\pa_\vf^2\r),
\end{eqnarray}
where the coefficient matrices can be obtained by straightforward
calculation. $(H_{\th\th}$, $H_{r\th}$, $H_{\th\vf}$, $H_{r\vf }$,
$H_{\vf\vf})$ are only functions of ($P_{kl}, M_{kl}, N_{kl}$), but
the others depend on their first order derivatives. The operator
$\pa_\vf$ should be replaced by matrix $D_\vf$ due to solution
(\ref{1.10}),
\begin{eqnarray}
\pa_\vf &\leftrightarrow& D_\vf=i~\diag (\kp,~ \kp+1,~ \kp,~ \kp+1),\\
\pa_\vf^2 &\leftrightarrow& D_\vf^2=-\diag \l(\kp^2, (\kp+1)^2,
\kp^2, (\kp+1)^2\r).
\end{eqnarray}
Then $[\hat H,\hat T]=0$ is equivalent to the following equations
\begin{eqnarray}
H_{\th\th}=H_{r\th} &=&
0,\label{3.9}\\
H_{\th}+H_{\th\vf}D_\vf &=&0,\label{3.10}\\
H_{r}+H_{r\vf }D_\vf &=&0,\label{3.11}\\
H_0+H_{\vf}D_\vf+H_{\vf\vf}D_\vf^2&=&0.\label{3.12}
\end{eqnarray}
The differential equations including in (\ref{3.9})-(\ref{3.12}) are
linear, and one component usually satisfies two independent
equations, which lead to constant solutions. So it is easy to
resolution.

By (\ref{3.9}), we can solve
\begin{eqnarray}
T_\th &=& \left( \begin {array}{cccc}  W_1 &-i W_2 e^{-i\phi}&
 W_3 & W_4 e^{-i\phi}
\\\noalign{\medskip} i W_2 e^{i\phi} & W_1
 & -W_4 e^{i\phi} & W_3 \\\noalign{\medskip} W_3 & -W_4
e^{-i\phi} & W_1
 & i W_2 e^{-i\phi}
\\\noalign{\medskip}W_4 e^{i\phi} & W_3
& -i W_2 e^{i\phi}& W_1 \end {array} \right),
\end{eqnarray}
in which $W_k=W_k(r,\th,\kp),~(k=1,2,3,4)$. Again by (\ref{3.10}),
we find that $W_1=W_3=W_4=0$, and $W_2\ne 0$ is a constant.
Obviously $\hat T$ can be only determined by a constant factor and
constant translation, so we choose $W_2=1$ to fix the solution.
Then in equivalent sense we have
\begin{eqnarray}
T_\th= K_\th. \end{eqnarray}

For the other equations in (\ref{3.10})-(\ref{3.12}), detailed
calculation shows that the solutions for $P_{kl}$ and $N_{kl}$ are
underdetermined. Expressing $P_{kl}$ by $N_{kl}$, we can get the
following results by simple calculation
\begin{eqnarray}
&V=V(r),\qquad A = r^{-1} U(\th),\qquad \pa_\th F_\ga=0,\qquad
F_\be =
0,&\label{3.15}\\
&T_0-\hbar i T_\vf D_\vf = \l( K_0-\hbar i K_\vf D_\vf \r)+ e U
\diag(\sg_\th,-\sg_\th)+\la_0,& \label{3.16}
\end{eqnarray}
where $\la_0$ is a constant. If we set $\hat T=\hat K$ when $U=0$,
we get $\la_0=0$. Substituting $D_\vf \to \pa_\vf$ into
(\ref{3.16}), in equivalent sense, we finally get
\begin{eqnarray}
T_0 = K_0+ e U \diag(\sg_\th,-\sg_\th),\qquad T_\th=K_\th,\qquad
T_\vf= K_\vf.\label{3.17}
\end{eqnarray}

In the spherical coordinate system, the above procedure is
invertible, so equations (\ref{3.15}) actually form the sufficient
and necessary conditions of the separation of variables to Dirac
equation with axisymmetry. Noticing the relation (\ref{3.17}),
conditions (\ref{3.15}) can also be verified by the commutative
relations (\ref{2.22})-(\ref{2.24}).

$\{V=A=F_\be=\pa_\th F_\ga=0\}$ is the simplest nonlinear case
discussed in \cite{4}-\cite{10}.

$\{V=V(r), A=F_\be=0, F_\ga=G(r)\}$ are the cases solved in
\cite{17} and \cite{18}-\cite{26}.

$\{V=\frac {Ze} r , A=\frac e {r\sin\th}(a-b\cos\th), F_\be=
F_\ga=0\}$ is the case solved in \cite{27}.

For the general case of (\ref{3.15}), although the separation of
variables is valid for the Dirac equation (\ref{1.5}), but the
solutions to the reduced ordinary differential equations usually can
not be expressed by elementary functions. So some auxiliary
numerical computation is still necessary.

For the Schr\"odinger equation
\begin{eqnarray}
\hbar i \pa_0\phi=\hat H\phi,  \quad \hat H\equiv -\frac
{\hbar^2}{2 m}\l(\pa_r^2+\frac 2 r \pa_r+\frac 1 {\hbar^2r^2}\hat
L^2\r)+V(r,\th),\label{3.18}
\end{eqnarray}
the calculation is simple. Obviously, for $\hat L_z=-\hbar
i\pa_\vf$, we have
\begin{eqnarray}
[\hat H,\pa_\vf]=0,\qquad \pa_\vf\leftrightarrow \kp
i.\label{3.19}
\end{eqnarray}
We look for the following second order operator
\begin{eqnarray}
\hat Y=Y_0+Y_\th \pa_\th+\hat L^2.\label{3.20}
\end{eqnarray}
By $[\hat Y,\pa_\vf]=0$, we get
\begin{eqnarray}
\pa_\vf Y_0=\pa_\vf Y_\th=0.\label{3.21}
\end{eqnarray}
By $[\hat H,\hat Y]=0$, we find
\begin{eqnarray}
V &=& W(r)+\frac 1 {r^2}U(\th),\label{3.22}\\
Y_0 &=& \la_0-{2m}U(\th),\qquad Y_\th =0,
\end{eqnarray}
where $U(\th)$ and $W(r)$ are arbitrary functions. (\ref{3.22}) is
the condition to solve the eigenfunctions of (\ref{3.18}) by
separating variables. Different from the Dirac equation,
(\ref{3.22}) shows $V$ can vary with $\th$.

\section{discussion and conclusion}
\setcounter{equation}{0}

From the above calculation, we get the following conclusion and
remarks:

\begin{enumerate}
\item  The above procedure has general significance for
separation of variables. The separation of variables is equivalent
to the existence of the following Hermitian operators chain
\begin{eqnarray}
\hat H_1=H_1(\pa_1),\quad \hat H_2=H_2(\pa_1,\pa_2),\quad
\cdots,\quad \hat H_n=H_n(\pa_1,\pa_2,\cdots,\pa_n),\label{4.2}
\end{eqnarray}
which form a Abelian Lie algebra
\begin{eqnarray}
[\hat H_j,\hat H_k]=0,\qquad (j,k=1,\cdots, n).\label{4.3}
\end{eqnarray}
(\ref{4.3}) forms the integrable conditions of the equation, and can
be obtained via simple calculation. In general, a complete chain
similarly to the case of (\ref{1.5}) may be absent, but the
incomplete chain is still helpful to simplify the dynamic equation.

\item   The commutative relations (\ref{4.3}) are conditional
relations, that is, it is enough to hold only for the
eigenfunctions, rather than hold in the whole function space of the
solutions. The solution to the nonlinear spinors is an
expample\cite{4}-\cite{10}. So we can use the information of
solutions for solving the operators chain.

\item  The above method is related to the concept
`superintegrability', which was introduced at the classical level by
Wojciechowski\cite{29} and at the quantum level by
Kuznetsov\cite{30}. Superintegrability also deals with the symmetry
and integrability of a Hamiltonian system, and many elaborate models
were solved\cite{29}-\cite{41}.
\end{enumerate}

\section*{Acknowledgments}
The author is grateful to his supervisor Prof. Ta-Tsien Li and Prof.
Han-Ji Shang for their encouragement. The work is supported by the
Postdoctoral Station of Fudan University.

\end{document}